\expandafter\def\csname ver@etex.sty\endcsname{3000/12/31}
\documentclass[colorlinks=true, linkcolor=blue, citecolor=blue, urlcolor=blue]{aa}  

\usepackage{orcidlink}
\usepackage{txfonts}
\usepackage{graphicx}	% Including figure files
\usepackage{amsmath}	% Advanced maths commands
\usepackage{amssymb}	% Extra maths symbols
\usepackage{xcolor}     % Text color
\usepackage{soul}       % for strikeout, can be removed after comments are treated 
\usepackage{array}
\usepackage{makecell}
\usepackage{hyperref}
\usepackage{multirow}

\defcitealias{Sheu_2024}{0}
\defcitealias{odonnell_carousel}{I}
\defcitealias{urcelay2026}{II}

\makeatletter
\newcommand{\mycustombox}[1]{%
  \fbox{%
    \begin{minipage}{\linewidth-2\fboxsep-2\fboxrule}
      \vspace{0.15cm} 
      
      \raggedleft 
      \textbf{Prepared for Submission} --- \textit{The Open Journal of Astrophysics}\\
      Draft version: \today
      
      \vspace{0.15cm}
    \end{minipage}%
  }%
}

\patchcmd{\aa@maketitle}{\fbox}{\mycustombox}{}{}
\makeatother

\makeatletter
\renewcommand*\aa@textidlineempty{\aa@headings}
\renewcommand*\aa@headfont{\small\mathversion{normal}}
\makeatother

\begin{document} 

\title{
The Carousel Lens III: Joint lens convergence and source kinematic modeling using MUSE integral-field unit data}
\author{William~Sheu\orcidlink{0000-0003-1889-0227}\inst{\ref{af:ucla}}\corrauth{wsheu@astro.ucla.edu} 
\and 
Xiaosheng~Huang\orcidlink{0000-0001-8156-0330}\inst{\ref{af:scu}, \ref{af:lbnl}}
\and
Tania~M.~Barone\orcidlink{0000-0002-2784-564X}\inst{\ref{af:Swinburne}, \ref{af:ASTROThreeD}, \ref{af:cfa}} 
\and 
Fuyan~Bian\orcidlink{0000-0002-1620-0897}\inst{\ref{af:ESOChile}} 
\and 
Aleksandar~Cikota\orcidlink{0000-0001-7101-9831}\inst{\ref{af:GeminiObs}} 
\and
Karl~Glazebrook\orcidlink{0000-0002-3254-9044}\inst{\ref{af:Swinburne}, \ref{af:ASTROThreeD}} 
\and
Tesla~Jeltema\orcidlink{0000-0001-6089-0365}\inst{\ref{af:ucsc}, \ref{af:scipp}} 
\and 
Evan~Odell\orcidlink{0009-0006-9722-2401}\inst{\ref{af:berkeley_astro}} 
\and 
Jackson~H.~O'Donnell\orcidlink{0000-0003-4083-1530}\inst{\ref{af:ucsc}, \ref{af:scipp}} 
\and 
Linus~Upson\orcidlink{0009-0003-6028-5677}\inst{\ref{af:berkeley_physics}} 
\and
Felipe~Urcelay\orcidlink{0009-0009-0407-2419}\inst{\ref{af:oxford}} 
\and 
Sean~Xu\orcidlink{0009-0006-4623-3629}\inst{\ref{af:berkeley_physics}, \ref{af:lbnl}} 
}

\institute{
Department of Physics and Astronomy, University of California, Los Angeles, CA 90095, USA \label{af:ucla} 
\and Department of Physics \& Engineering Physics, Santa Clara University, Santa Clara, CA 95053, USA \label{af:scu}
\and Physics Division, Lawrence Berkeley National Laboratory, Berkeley, CA 94720, USA \label{af:lbnl}
\and Centre for Astrophysics and Supercomputing, Swinburne University of Technology, Hawthorn, Victoria 3122, Australia \label{af:Swinburne}
\and ARC Centre of Excellence for All Sky Astrophysics in 3 Dimensions (ASTRO 3D), Australia \label{af:ASTROThreeD}
\and Center for Astrophysics, Harvard \& Smithsonian, Cambridge, MA 02138, USA \label{af:cfa}
\and European Southern Observatory, Alonso de C\'ordova 3107, Vitacura, Santiago, Chile \label{af:ESOChile}
\and Gemini Observatory, NSF's NOIRLab, La Serena, Chile \label{af:GeminiObs}
\and University of California, Santa Cruz, Santa Cruz, CA 95064, USA \label{af:ucsc}
\and Santa Cruz Institute for Particle Physics, Santa Cruz, CA 95064, USA \label{af:scipp}
\and Department of Astronomy, University of California, Berkeley, CA 94720, USA \label{af:berkeley_astro}
\and Department of Physics, University of California, Berkeley, CA 94720, USA \label{af:berkeley_physics}
\and Sub-department of Astrophysics, University of Oxford, Oxford OX1 3RH, UK \label{af:oxford}
}

%\datesmooth{\today}
\abstract{
We introduce a new, independent modeling component to the Carousel Lens series by exploiting the spatially-resolved kinematics contained within the lensed arcs, which so far has remained rather underexplored in strong-lensing analyses. We present an end-to-end forward-modeling framework that reconstructs an integral-field spectroscopic datacube from a parameterization of the lens convergence profile together with the line-of-sight velocity, velocity dispersion, and surface brightness profiles of the unlensed source. The kinematic structure of the source acts as an additional set of ray-traceable markers distributed across every lensed image, inaccessible from imaging data alone. Furthermore, by generating an [\ion{O}{II}] emission-line datacube, we isolate a single source redshift and remove lens and line-of-sight light contamination. We apply this framework to deep VLT/MUSE observations of the Carousel Lens, modeling the [\ion{O}{II}] emission of the quadruply-imaged star-forming galaxy at $z_{\rm s} = 1.432$ that anchors the system's cosmography constraints. We measure an Einstein radius $\theta_{\rm E} = 12\farcs816^{+0.005}_{-0.007}$ and a logarithmic slope $\gamma = 1.412^{+0.002}_{-0.002}$, tightening the statistical uncertainties on both by roughly a factor of $2-2.5$ relative to the corresponding flat emission-line map model. We find that the source galaxy exhibits a strong rotation, lies confidently along the star-forming main sequence, and has an inferred central black hole mass of $\log M_{\rm BH}/M_{\odot} = 7.89 \pm 0.51$. These results establish datacube-level forward modeling as a practical and independent route to traditional image-based lens models in the current systematics-limited regime, and one that extends naturally to this system's other lensed sources and to other lens systems.
}


\titlerunning{The Carousel Lens III: Joint lens convergence and source kinematic modeling}
\authorrunning{W. Sheu et al.}
\maketitle
\nolinenumbers

%-------------------------------------------------------------------
\section{Introduction} \label{sec:introduction}
Strong gravitational lensing is a leading probe in observational cosmology.  Measuring the time delays of variable sources allows for an independent late-time measurement of the Hubble constant, with the potential precision to break the Hubble tension \citep[e.g.,][]{tdcosmo2025, Pascale2025, sheu1433, galan2026}.  By using the properties of compound lenses (i.e., lenses with multiple sources at different redshifts), it is also possible to constrain the dark energy equation-of-state and the matter density of the universe \citep[e.g.,][]{Caminha_2022, urcelay2026, ballard2026}.  By measuring flux and time-delay anomalies for lensed quasars, one can also assess the dark matter fraction and substructure \citep[e.g.,][]{Xu_2015, Gilman2024, Paugnat_2025}.  And perhaps in the most direct application, strong lensing allows us to study magnified, high-redshift sources \citep[e.g.,][]{Treu_2022}.  All these applications rely on being able to model the lens mass profile precisely and accurately.  In other words, tighter constraints on these lens models would propagate as tighter constraints on all these fields of cosmology.  

\begin{figure*}
\begin{center}
 \includegraphics[width=1\linewidth]{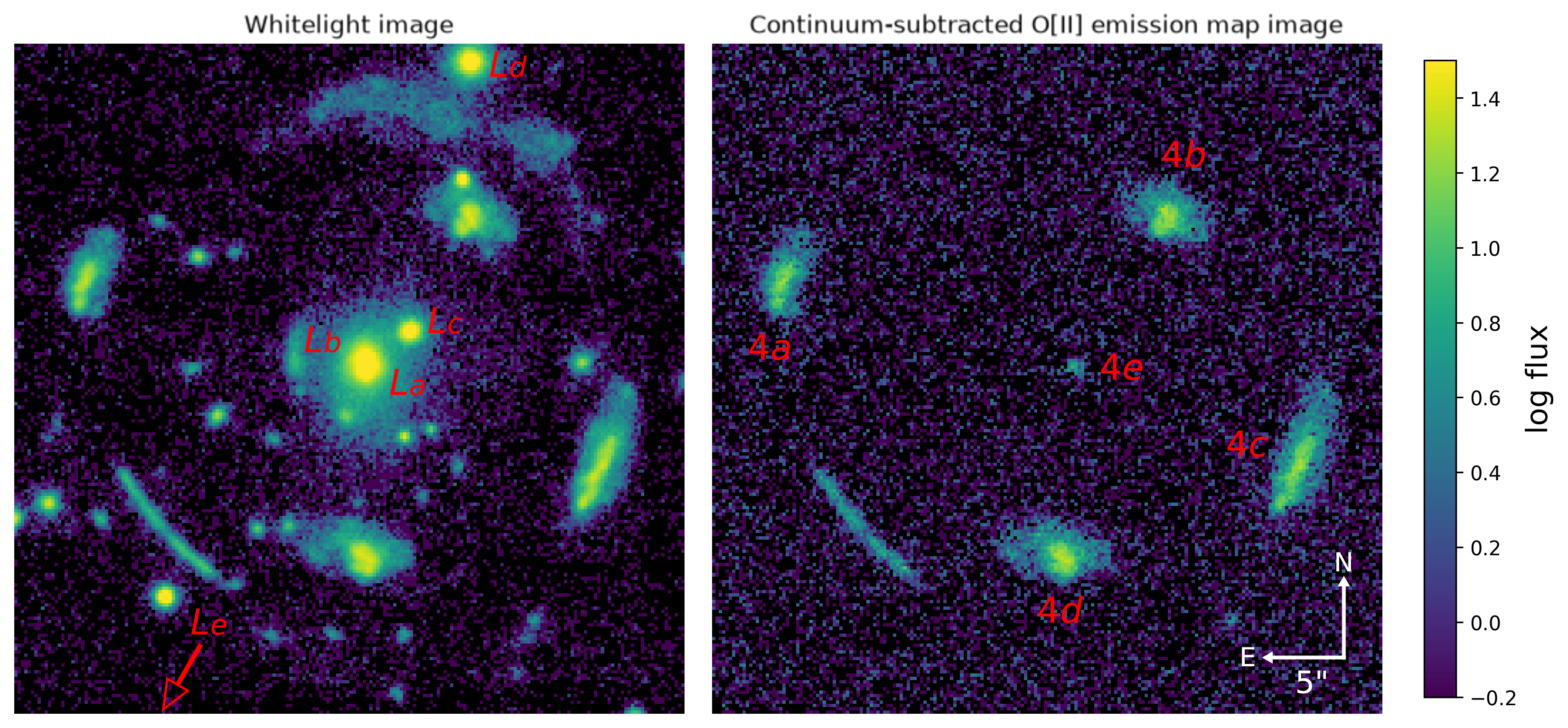}
 \caption{VLT/MUSE images of the Carousel Lens.  Left: the full whitelight image, generated by aggregating the datacube along its spectral axis.  The most-prominent galaxy-cluster members are labeled here, with $L_{\rm e}$ lying outside the image cutout.  This is similar to what is typically observed with imaging data.  Right: by focusing on the [\ion{O}{II}] emission at $z_{\rm s}=1.432$ and leveraging the continuum of the other sources in frame, we can generate an image that isolates the source-of-interest S4.  Note that source galaxy 5 lies at the same redshift as S4, hence the extra arc to the southeast and image to the southwest within the cutout; we simply mask out these images in our modeling scheme.  We also introduce the same labeling of the S4 lensed images as the Carousel Lens Papers~\citetalias{Sheu_2024}, \citetalias{odonnell_carousel}, and \citetalias{urcelay2026}.}
 \label{fig:wl_comparision_plot}
\end{center}
\end{figure*}

These lens models are usually constructed primarily with imaging data, using the lensed-source positions and arcs as constraints in the modeling pipeline.  More accurate models also use spectroscopic data of the lens galaxy (aside from redshift, such as the internal kinematic information) to place additional constraints on the lens galaxy mass \citep[e.g.,][]{Sheu_dinos, Knabel_2025}.  While this can be done with slit and aperture spectroscopy, the most robust models use spatially-resolved spectroscopy from integral-field unit (IFU) spectrographs.  Being able to probe the full kinematic map of the lens galaxy allows for not only constraints on the total mass, but also on the slope of the lensing profile.  %Aside from this, IFU data has been used (in the context of strong lensing) to probe the emission lines and continuum of lensed quasars \citep[e.g.,][]{Paugnat2026}, and for source redshift identification \citep[e.g.,][]{Sheu_2024, lin2026, odonnell_carousel}.

Despite these applications, we posit that strong-lensing IFU information is still underutilized.  Rather than providing external constraints to the lens modeling pipeline, the idea of using the spatially-resolved kinematic information within the arcs themselves is rarely pursued.  This methodology uses the kinematic markers within the source galaxy as lens modeling constraints, introducing additional velocity dimensionalities to the process (in our manuscript, a ``marker'' is defined as an identified feature on the lensed images that should ray-trace to the same location on the source).  In addition to relying on high-resolution imaging to identify flux-based markers within each lensed image, leveraging this extra dimension can achieve the same effect with kinematics-based markers.  Additionally, IFU data can be used to isolate light from a single source galaxy redshift, often removing nearly all lens and field galaxy light contamination.  This 
significantly reduces, if not eliminates,  degeneracies between the lens and source light, and crucially, reveals radial and demagnified central images overwhelmed by the lens galaxy in imaging data.  These central images provide tight constraints on the lensing mass slope near the cluster core, well within the tangential critical curves, that are challenging to obtain in other ways, enabling critical tests of the cold dark matter model and alternatives such as the self-interacting dark matter \citep{Sagunski2021, andrade2022, odonnell2026}.  Lastly, by forward-modeling the lens profile with the unlensed kinematic profile of the source, we can build a precise dynamical map of the source, incorporating data from all the lensed images.  Having such a precise model makes it possible to even probe the black hole mass within the source galaxy, which would not be possible with imaging data alone \citep{Newman_2026}.

Here, we introduce a novel, spatially-resolved spectroscopy modeling pipeline to incorporate the dynamical information of the source into a macro model of the lensing system.  Previous iterations of this idea \citep{Bolton_2007, Rizzo_2018, Chiriv_2020, Zhou2025} differ in that we utilize an end-to-end forward modeling framework to simulate each individual spaxel (spatial pixel of the IFU) from a parameterization of the lens model and the source kinematics (the line-of-sight velocity, velocity dispersion, and surface-brightness profile).  Our pipeline was built with the intention of applying it to real data, which we do in this paper using the Carousel Lens with a quadruply lensed late-type galaxy at $z_{\rm s}=1.432$.  

% In \citet{Bolton_2007}, the technology at the time was not at the level to fully utilize the methodology.  However, with high spacial-resolution IFU instruments such as MUSE, Keck KCWI, and \textit{JWST} NIRSPEC, this technique can reach its fullest potential.  

This paper is organized as follows. Section~\ref{sec:carousel} introduces the Carousel Lens and the previous papers in this series. Section~\ref{sec:muse} describes the Multi Unit Spectroscopic Explorer (MUSE) observations of the Carousel Lens and data reduction. Section~\ref{sec:methodology} details the datacube reconstruction and modeling framework. Section~\ref{sec:results} presents the application to the Carousel Lens, highlighting the improved constraints on the mass model and the recovered source dynamics. Finally, we conclude in Section~\ref{sec:conclusion}.

\section{The Carousel Lens} \label{sec:carousel}
\begin{figure*}
\begin{center}
 \includegraphics[width=0.85\linewidth]{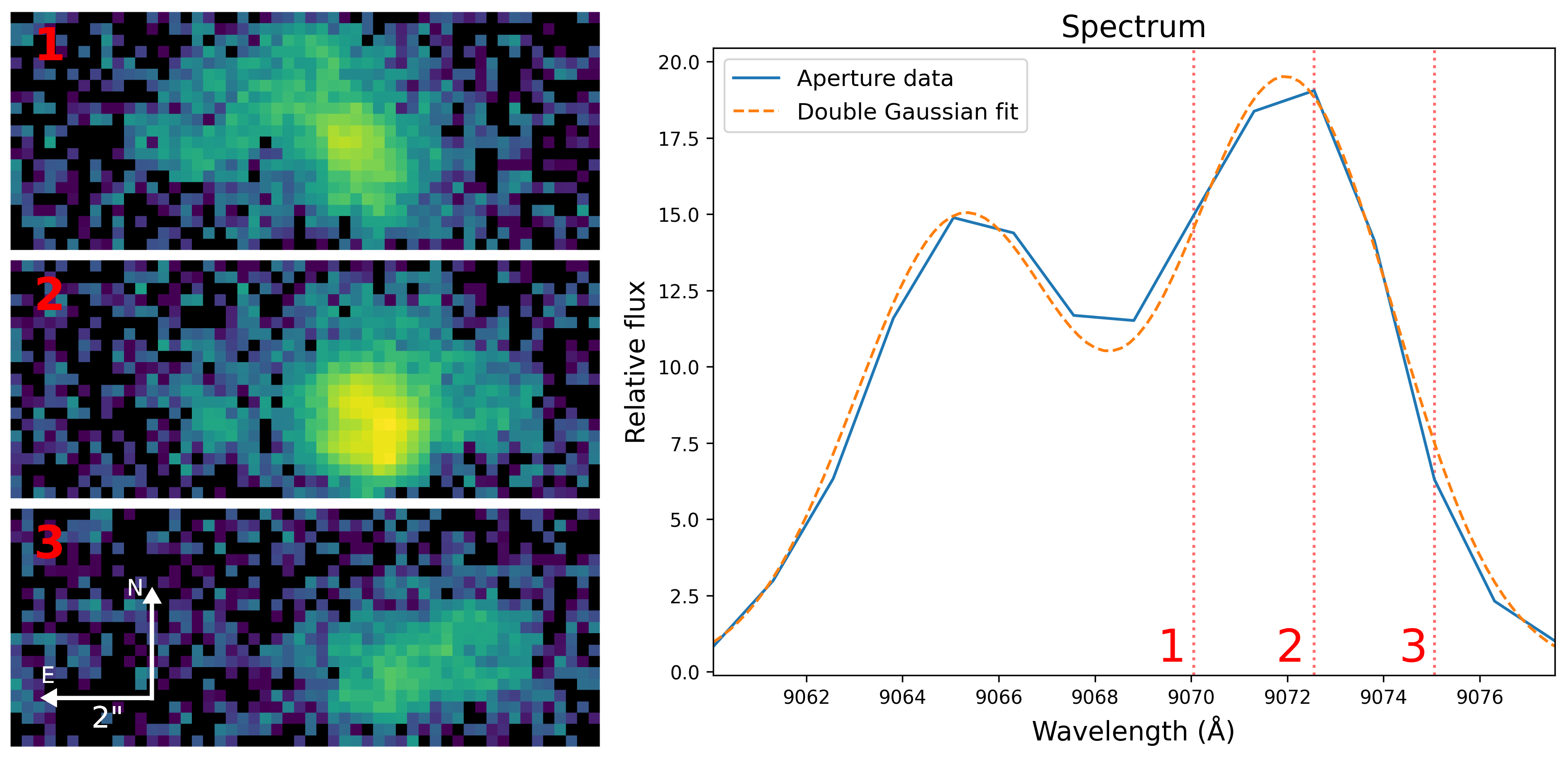}
 \caption{A static presentation of the continuum-subtracted datacube, specifically image $4d$ (the southern-most image in Figure~\ref{fig:wl_comparision_plot}).  Left panels: the image at different spectral slices of the datacube.  Right panel: the large aperture spectrum of S4, plotted alongside the double Gaussian fit in our datacube modeling.  The red dotted lines indicate the wavelength slices 1 (9070.06\AA), 2 (9072.56\AA), and 3 (9075.06\AA) to which the left panel slices correspond.  From the left panels, we clearly see that the emission of S4 shifts from east to west with increasing wavelength, indicating a strong rotation present.  For a full non-static presentation of the datacube, see here: \url{https://portal.nersc.gov/cfs/deepsrch/IFUL/figure2.gif}.}
 \label{fig:wave}
\end{center}
\end{figure*}

The Carousel Lens is a remarkably relaxed cluster-scale lens, independently discovered by multiple groups \citep{jacobs2017, huang2021, odonnell2022a}.  The cluster BCG lies at a redshift of $z_{\rm l} = 0.488$, with multiple lensed arcs from $0.962 \le z_{\rm s} \le 4.090$ \citep[][Carousel Lens Paper~0 and I, respectively]{Sheu_2024, odonnell_carousel}.  Despite being at a cluster scale, what makes this system particularly valuable for cosmology is its relaxed nature, with all lensed arcs displaying textbook galaxy-scale lensing configurations.  In Paper~0 \citep{Sheu_2024}, an initial model from \textit{Hubble Space Telescope} (\textit{HST}) imaging was constructed using the primary sources spectroscopically identified by shallow Very Large Telescope (VLT) MUSE IFU data.  The most prominent and best-constrained lensed feature, dubbed source family 4, is a quadruply lensed star-forming galaxy, in an ``Einstein cross'' configuration, with an Einstein radius of $\theta_{\rm E} \approx 13''$ and a redshift of $z_{\rm s} = 1.432$.  From the initial MUSE data, it was remarked that source galaxy 4 (S4) exhibited significant LOS rotation.  

After obtaining deeper MUSE data, Paper~I \citep{odonnell_carousel} conducted a deep spectroscopic survey of the Carousel Lens environment, where they were able to reveal new sources and new images of known sources in the IFU data that were undetected in the \textit{HST} imaging data.  This includes the confirmation of lensed arcs at $z_{\rm s} > 3$.  The authors also find that there is a central demagnified image visible in the MUSE data after isolating the flux contribution from S4's redshift.  When modeled, this central image can be extremely useful in constraining the logarithmic slope of the lensing mass profile $\gamma$ at the core.  By isolating the flux to S4's [\ion{O}{II}] emission lines, they are able to disentangle S4 from the neighboring source 6 arcs and a foreground galaxy that lies directly on one of the lensed images.

With the additional spectroscopic information, a more refined lens model from the \textit{HST} imaging was constructed in Paper~II \citep{urcelay2026} and the first constraints on cosmology were made from the \textit{HST} arcs.  These constraints rely on measuring the Einstein radius for the different source redshifts.  The relation of the Einstein radii ratios versus redshift depends on cosmological parameters such as $\Omega_{\rm m}$ and $w$ (but is markedly independent of $H_{\rm 0}$).  Therefore, the precision of lens model measurements and especially $\theta_{\rm E}$ directly influence the constraints on cosmology.  Given the well-known covariance between the Einstein radius and the logarithmic slope $\gamma$, as well as the correlation between $\gamma$ and the deflection ratio in compound lenses, the precise and accurate measurement of $\gamma$ thus is also important for cosmology.  Furthermore, as S4's images are the most prominent lensing feature of the Carousel Lens, S4's redshift and its associated Einstein radius are used as the baseline for their $\Omega_{\rm m}$ and $w$ constraints in Paper~\citetalias{urcelay2026}, highlighting the importance of understanding the lens model derived from source family 4 alone.

\section{Integral field spectroscopy} \label{sec:muse}
Deep VLT/MUSE observations of the Carousel Lens system were conducted between December 2024 and January 2025 \citep[][Prog. ID 0114.A-2018A, PI: Barone]{2010SPIE.7735E..08B}. Utilizing the Wide-Field Mode (WFM) with adaptive optics (AO), we mapped a 1.28 square arcminute area using two overlapping pointings.  Eight 620-second exposures were obtained per pointing, totaling 2.76 hours of integration. Data reduction and coaddition into a single datacube were performed using the MUSE pipeline described in \citet{Weilbacher_2020}, and ZAP \citep{Soto_2016} was applied to further improve sky subtraction residuals.  The final datacube has a wavelength separation of 1.25\AA.  See Paper~\citetalias{odonnell_carousel} for more details on the reduction of the MUSE data, and for how it was used to spectroscopically analyze the Carousel Lens field.

A known problem with the MUSE data reduction pipeline is that it significantly underestimates the datacube covariance between spaxels \citep{Weilbacher_2020, Weilbacher_2022, Emsellem_2022}.  To compensate for this, we uniformly inflated the uncertainties by the subsampling factor squared, which conservatively overestimates our uncertainties relative to a treatment based on the full covariance matrix.

\begin{figure*}
\begin{center}
 \includegraphics[width=1\linewidth]{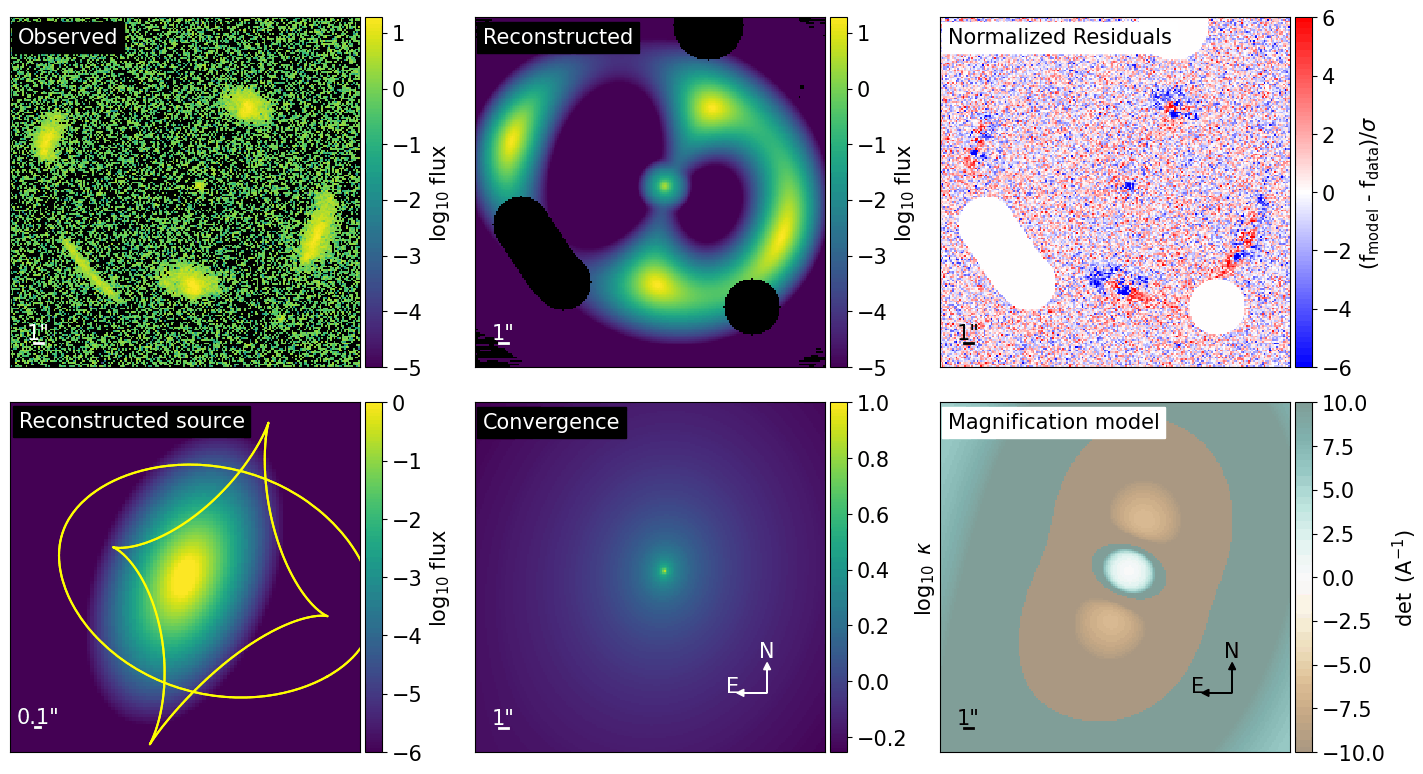}
 \caption{The initial, best-fitting flat model using \texttt{lenstronomy} assuming a single Sérsic light profile for the source galaxy, using the continuum-subtracted emission line map.  This fit serves as an initial parameterization for our datacube reconstruction.  See Section~\ref{subsec:2dmodeling} for the full model design.  Top left: the continuum-subtracted emission line map.  Top middle: the reconstructed \texttt{lenstronomy} image.  Top right: the normalized residuals between the two.  Bottom left: the source light profile on the lens plane, with the lensing caustics overlaid in yellow.  Bottom middle: the modeled convergence profile.  Bottom right: the image-plane magnification profile.  As expected, there are strong residuals due to the simplicity of the modeled source galaxy light profile, but nevertheless the image positions are well recovered.}
 \label{fig:lenstronomy_flat}
\end{center}
\end{figure*}

For our kinematic analysis of source family 4, we isolate its [\ion{O}{II}] emission lines and restrict our wavelength bounds to 9060.0$-$9077.5\AA\ (15 wavelength slices).  The point spread function (PSF) is probed using four isolated stars spectroscopically-identified within this truncated datacube, and well modeled using three concentric Gaussian functions.  We then subtract out the continuum by probing 12.5\AA\ above and below the wavelength bounds, measuring the median continuum flux in the two wavelength ranges, assuming a linear trend between them, and removing the interpolated continuum within our wavelength bounds.  This results in a datacube that isolates the flux from source families 4 and 5 (both of which lie at $z_{\rm s} = 1.432$).  For our narrow wavelength range, we expect the spectral resolution of $R=3500$ to be nearly constant across the data.  See Figure~\ref{fig:wl_comparision_plot} for a comparison between the whitelight image and the continuum-sub[\ion{O}{II}]tracted [OII] emission line map.  Also see Figure~\ref{fig:wave} for a visualization of the LOS rotation of S4, by honing in on specific wavelength slices of the datacube. 

\section{Methodology} \label{sec:methodology}
\begin{figure*}
\begin{center}
 \includegraphics[width=0.95\linewidth]{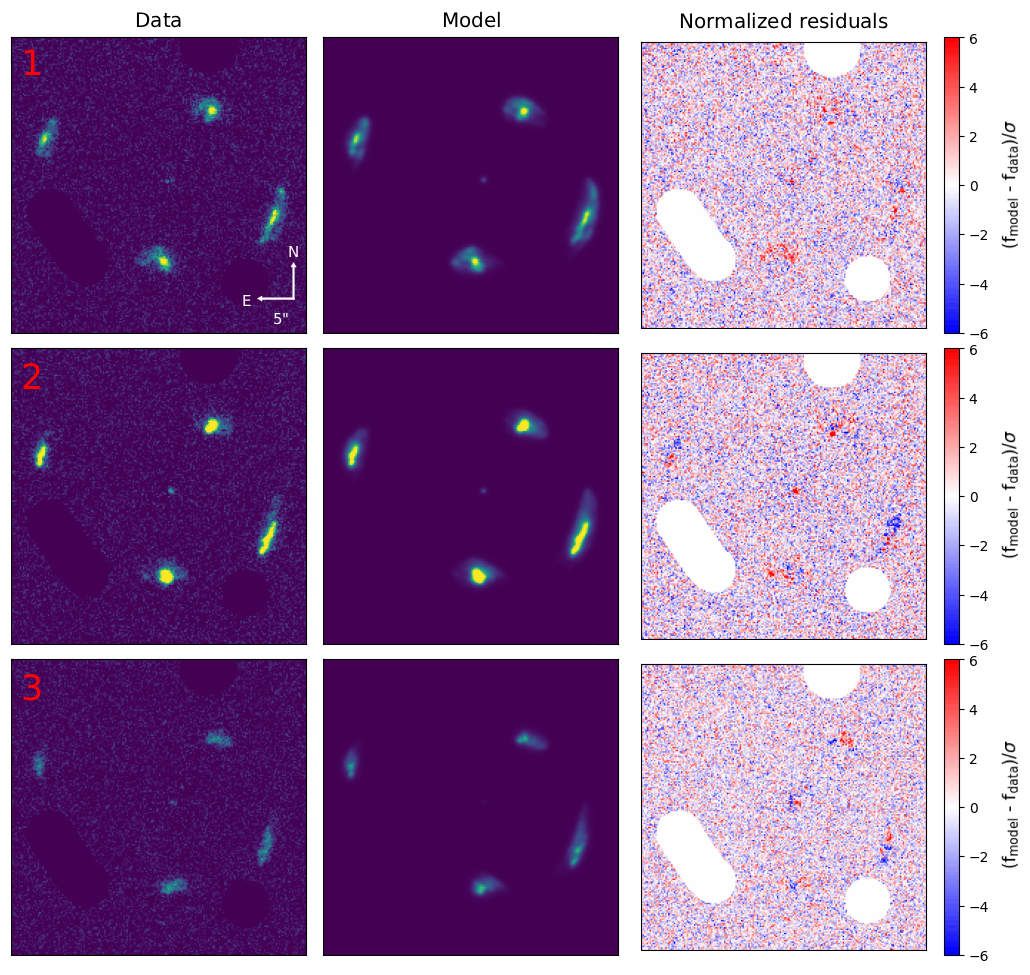}
 \caption{A static presentation of the best-fit datacube model and its comparison to the continuum-subtracted datacube.  Each row represents a specific wavelength slice of the datacube, with the wavelength labeling provided in Figure~\ref{fig:wave}.  Left: the continuum-subtracted datacube slice at the indicated wavelength.  Middle: the corresponding modeled datacube slice.  Right: the normalized residuals between the two.  For a full non-static presentation of the best-fitting datacube model, see here: \url{https://portal.nersc.gov/cfs/deepsrch/IFUL/figure4.gif}. }
 \label{fig:model_wave}
\end{center}
\end{figure*}

\begin{table*}
\begingroup
\renewcommand{\arraystretch}{1.25}
\begin{center}
\caption{The resulting posterior distributions for a subset of lensing parameters shared by the different modeling schemes.  The three models are (1) a flat, image-based model using \texttt{lenstronomy} with a single Sérsic source, (2) a flat, image-based model using \texttt{lenstronomy} with a single Sérsic source and an $n_{\rm max}=4$ shapelets basis, and (3) a datacube-based model using \texttt{IFUL} (see Section~\ref{sec:methodology} for the full details).  All angle quantities are given as East of North.  Constraints are significantly tighter across the common parameters when incorporating the source kinematic information into the lens model. }\label{tab:tam_comp}
\begin{tabular}{cc|ccc}
\hline
 Parameter & Units & Flat (Sérsic) & Flat (Sérsic + Shapelets) & \texttt{IFUL} (Kinematics) \\
 &  & (1) & (2) & (3) \\
\hline
$\theta_{\rm E}$ & [$''$] & $12.863^{+0.011}_{-0.012}$ & $12.816^{+0.011}_{-0.011}$ & $12.816^{+0.005}_{-0.007}$\\
$\gamma$ & & $1.431^{+0.005}_{-0.005}$ & $1.416^{+0.005}_{-0.005}$ & $1.412^{+0.002}_{-0.002}$\\
$q_{\rm m}$ & & $0.826^{+0.003}_{-0.003}$ & $0.837^{+0.003}_{-0.003}$ & $0.843^{+0.002}_{-0.001}$ \\
$\phi_{\rm m}$ & [$^\circ$] & $-81.49^{+0.31}_{-0.33}$ & $-81.12^{+0.32}_{-0.30}$ & $-81.52^{+0.18}_{-0.19}$\\
$\gamma_{\rm ext}$ & & $0.0193^{+0.0006}_{-0.0006}$ & $0.0180^{+0.0007}_{-0.0006}$ & $0.0191^{+0.0004}_{-0.0004}$\\
% $R_{\rm s}$ & [$''$] & $0.278^{+0.004}_{-0.004}$ & $0.278^{+0.012}_{-0.010}$ & $0.229^{+0.001}_{-0.002}$\\
$q_{\rm s}$ & & $0.572^{+0.004}_{-0.004}$ & $0.772^{+0.007}_{-0.007}$ & $0.696^{+0.002}_{-0.002}$\\
$\phi_{\rm s}$ & [$^\circ$] & $-68.93^{+0.32}_{-0.30}$ & $-71.91^{+1.68}_{-1.65}$ & $-62.36^{+0.12}_{-0.08}$\\
\hline
\end{tabular}
\end{center}
\endgroup
\end{table*}

In this section, we lay out our 3D datacube modeling framework, consisting of an initial 2D image modeling (Section~\ref{subsec:2dmodeling}), source light binning (Section~\ref{subsec:binning}), and the datacube reconstruction  (Section~\ref{subsec:datacubemodeling}).

\subsection{2D image modeling} \label{subsec:2dmodeling}
Prior to fully utilizing the spectral dimension of the datacube, we first use the flat, emission line image to find an initial parameterization of the model.  We use the lens modeling software \texttt{lenstronomy}\footnote{\url{https://lenstronomy.readthedocs.io/en/latest}} \citep{birrer2018, lenstronomyII} to build this initial model.  \texttt{lenstronomy} has been shown to be robust in the Time-Delay Lens Modeling Challenge \citep[TDLMC;][]{Ding2021}, where two independent teams achieved consistency within Rung 2.  As previous papers in this series have shown the Carousel Lens to be extremely relaxed, we find that a single elliptical power-law profile with an external shear component well describes the primary convergence of the cluster.  As shown in Papers~\citetalias{Sheu_2024} and~\citetalias{urcelay2026}, this assumption holds true across a range of source redshifts and Einstein radii; this should hold especially true in our analysis, as we are looking at only a single redshift source.  We experimented with adding a profile to cluster members $L_{\rm d}$ and $L_{\rm e}$ (see Figure~\ref{fig:wl_comparision_plot} for the labeling scheme), but we find that the S4 images do not constrain $L_{\rm d}$ at all, with the model strongly preferring an $L_{\rm d}$ Einstein radius of zero, despite adding a prior on it based on its velocity dispersion.  For this reason, we choose to not include $L_{\rm d}$ in our model, and to add a fixed singular isothermal sphere (SIS) model to $L_{\rm e}$.  For $L_{\rm e}$, we fix $\theta_{\rm E} = 1\farcs59$, derived from the measured aperture velocity dispersion of $\sigma_{v} = 313$~km~s$^{-1}$ from the MUSE data and the assumption of an SIS profile:
\begin{equation}
    \theta_{\rm E} \approx 4\pi\frac{\sigma^2_{v}}{c^2}\frac{D_{\rm ls}}{D_{\rm s}}.
\end{equation}
Our measurement of the Einstein radius from the kinematics is very close to what was measured through lens modeling in Paper~\citetalias{urcelay2026} ($\theta_{\rm E} = 1\farcs68$).  We also fix the location of the SIS to the brightest pixel of $L_{\rm e}$.  Given that $L_{\rm e}$ is $\sim40''$ away from the brightest cluster galaxy (BCG), we determine these modeling assumptions to be necessary for a well-constrained fit using only S4 information.  Therefore, our lensing profile consists of an elliptical power law as our primary lens with its Einstein radius ($\theta_{\rm E}$), logarithmic slope ($\gamma$), ellipticity (semi-major and minor axes ratio $q_{\rm m}$ and position angle $\phi_{\rm m}$), and centroid ($c_{\rm x}$, $c_{\rm y}$) being fitted, a fixed SIS profile at $L_{\rm e}$, and an external shear component (shear strength $\gamma_{\rm ext}$ and directionality $\phi_{\rm ext}$).

As there is no lens light to model, we only have to account for the source light profile, for which we use a single elliptical Sérsic profile \citep{sersic} to describe S4, parameterized by its half-light radius ($R_{\rm s}$), Sérsic index ($n$), ellipticity (semi-major and minor axes ratio $q_{\rm s}$ and position angle $\phi_{\rm s}$), and centroid ($c_{\rm s, x}$, $c_{\rm s, y}$).
We will introduce additional complexity to the source light profile in the later steps, as this 2D image modeling step is to generate a suitable initial parameterization.  

Therefore, we fit a total of 14 parameters in our emission-line map image modeling: eight parameters describing the convergence profile \{$\theta_{\rm E}$, $\gamma$, $q_{\rm m}$, $\phi_{\rm m}$, $c_{\rm x}$, $c_{\rm y}$, $\gamma_{\rm ext}$, $\phi_{\rm ext}$\} and six describing the source galaxy light \{$R_{\rm s}$, $n$, $q_{\rm s}$, $\phi_{\rm s}$, $c_{\rm s, x}$, $c_{\rm s, y}$\}.  We run a particle swarm optimization \citep[PSO;][]{pso} algorithm to probe the parameter space, and then we run a Markov Chain Monte Carlo (MCMC) algorithm with \texttt{EMCEE} \citep{emcee} until convergence to measure the final parameter posteriors.  We use the best-fitting parameterization (present in Figure~\ref{fig:lenstronomy_flat}) as a starting point for the subsequent steps.

\subsection{Source light binning} \label{subsec:binning}
\begin{figure*}
\begin{center}
 \includegraphics[width=1\linewidth]{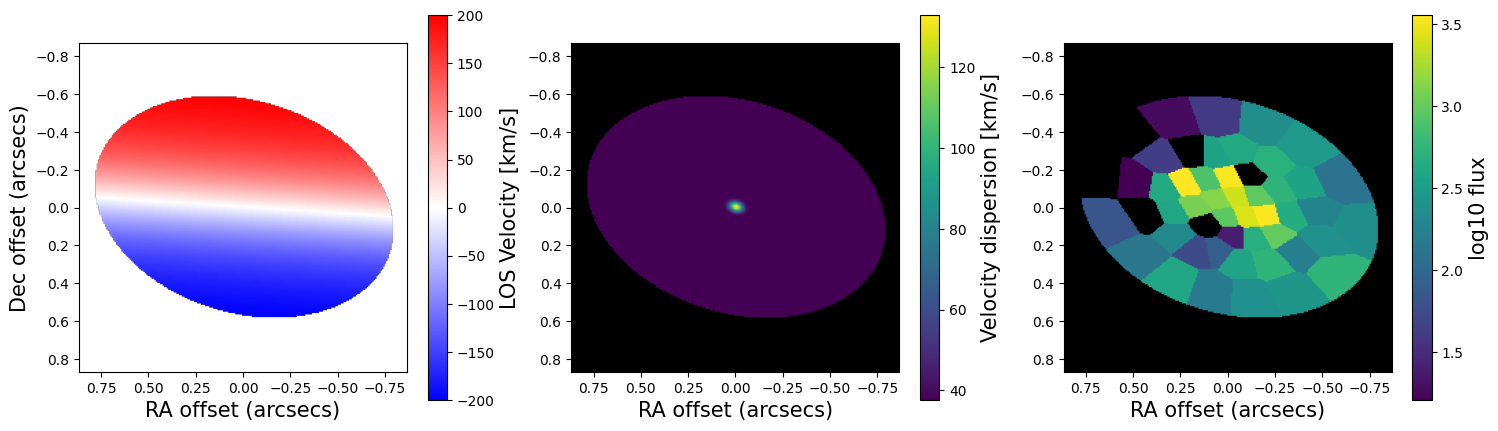}
 \caption{The best-fit source-plane kinematics of S4.  Left: the line-of-sight velocity $v_{\rm LOS}$ profile.  Middle: the velocity dispersion $\sigma_{v}$ profile.  Right: the discretized logarithmic source surface-brightness profile.  }
 \label{fig:source_kinematics}
\end{center}
\end{figure*}

In this step, we lay the architecture for adding details to the source galaxy, by discretizing the light profile into bins.  To do this, we use the \texttt{PowerBin}\footnote{\url{https://pypi.org/project/powerbin}} software to bin the unlensed light on the source plane \citep{Cappellari2025}.  Up to an elliptical radius of $3R_{\rm s}$, we generate exactly 50 bins of approximately equivalent signal from the modeled Sérsic profile (see the right-panel of Figure~\ref{fig:source_kinematics} for the bin boundaries).  

\texttt{PowerBin} is typically applied to bin observed data with a set pixel size.  However, as there is no physical grid size on the source plane, additional processing must be applied such that this discretization remains robust against any sampling of points on the source plane.  We calculate the \texttt{PowerBin} bin centroids, and parameterize them by their elliptical polar coordinates, relative to the Sérsic profile center and ellipticity.  We then determine the bin group for any arbitrary point on the source plane by solving for which bin center is closest (excluding points outside of the $3R_{\rm s}$ elliptical boundary).  Through this parametrization, any given source-plane point within the $3R_{\rm s}$ region will be categorized into a corresponding bin.  Additionally, since we define the bin centroids on an elliptical polar coordinate system based on the Sérsic profile, this allows us to flexibly change and fit the overall size and shape of the source galaxy light without regenerating bins between each iteration of the datacube modeling.

\begin{figure}
\begin{center}
 \includegraphics[width=1\linewidth]{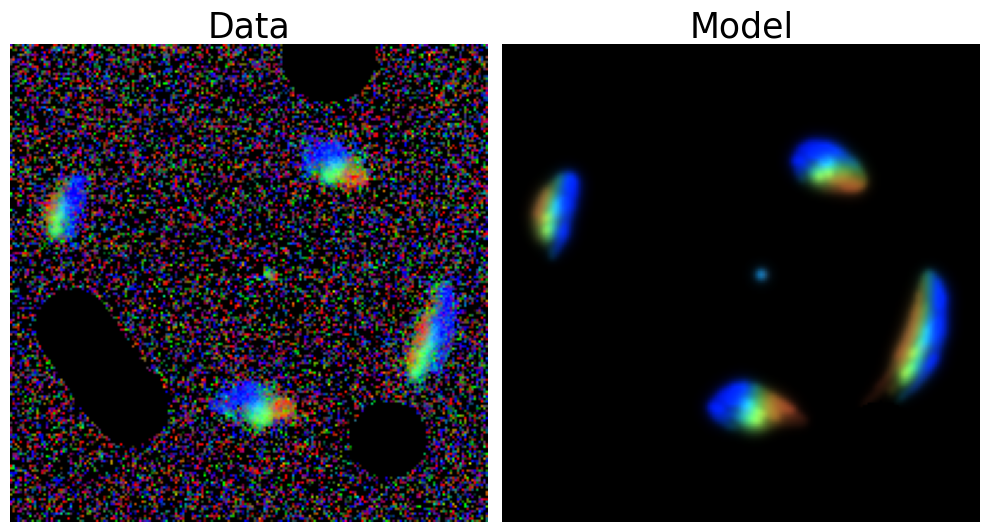}
 \caption{RGB image representations of the continuum-subtracted and the modeled datacubes.  The red, green, and blue channels correspond to the wavelength slices 1 (9070.06\AA), 2 (9072.56\AA), and 3 (9075.06\AA) in Figures~\ref{fig:wave} and \ref{fig:model_wave}.}
 \label{fig:rgb_comp}
\end{center}
\end{figure}

\subsection{Datacube reconstruction}\label{subsec:datacubemodeling}
At any given point of the unlensed image of the source galaxy, there are three primary properties that will affect the datacube: the surface-brightness, the line-of-sight velocity $v_{\rm LOS}$, and the line-of-sight velocity dispersion $\sigma_{v}$ (ignoring higher order kinematic effects).  Due to computational limitations, we are currently unable to build an all-encompassing dynamical model in conjunction with the lens modeling scheme.  Hence, in order to reconstruct the datacube, we parametrically describe all three profiles of S4.

As we have already identified clear rotation within S4, we choose to describe the $v_{\rm LOS}$ with the arc-tangent function:
\begin{equation}\label{eq:vlos}
    v_{\rm LOS}(r_{\rm line}) = \frac{2}{\pi}A_{v} \arctan \left(\frac{r_{\rm line}}{R_{v}}\right) + C_{v},
\end{equation}
where $A_{v}$, $R_{v}$, and $C_{v}$ are parameters we constrain in our datacube modeling \citep{Courteau1997, Pizagno2007, Bloom_2017, Tiley2019}.  $A_{v}$ is the asymptotic rotation velocity, $R_{v}$ is the rotational scale radius, and $C_{v}$ is redshift velocity.  $r_{\rm line}$ is defined as the distance from a line that crosses through the centroid of the source galaxy (parameterized by its position angle ${\phi}_{v}$).  And so, there are four parameters that describe the LOS velocity profile of S4: \{$A_{v}$, $R_{v}$, $C_{v}$, ${\phi}_{v}$\}.

From the rotation, complex light distribution, and strong emission lines, we can safely assume that S4 is a late-type, star-forming disk galaxy.  As such, we adopt a line-of-sight velocity dispersion model consisting of a uniform intrinsic velocity dispersion floor added in quadrature to a Gaussian central shear peak:
\begin{equation}\label{eq:sigmav}
    \sigma_{v}(r) = \sqrt{A_{\sigma}^2 \exp\left(-\frac{r^2}{2 R_{\sigma}^2}\right) + C_\sigma^2}
\end{equation}
Here, $C_{\sigma}$ represents the turbulent gas dispersion floor characteristic of gas-rich disks at $z \sim 1.5$ \citep[e.g.,][]{Schreiber2009, Wisnioski_2019}, while the Gaussian component accounts for the unresolved inner rotational velocity gradient smeared across the effective resolution element \citep[e.g.,][]{Epinat2012}. The spatial scale $R_\sigma$ reflects the combination of the galaxy's intrinsic turnover radius and the MUSE beam projected onto the kinematic axis, and $A_\sigma$ represents bulge velocity dispersion of the core.  Thus, there are three parameters (\{$A_\sigma$, $R_\sigma$, $C_\sigma$\}) that describe our model $\sigma_{v}$ profile.

Finally, for the surface-brightness profile, we make use of the bins described in Section~\ref{subsec:binning}.  That is, for each binned region, a constant independent surface-brightness amplitude is parameterized.  Much like in \texttt{lenstronomy}, each bin amplitude is linearly fitted at each iteration of the model, allowing us to omit directly sampling these parameters and avoid drastically increasing our parameter space.  If a negative surface brightness is derived from this method, it is scaled up to zero.  As discussed in the previous section, we still sample macro-quantities of the light profile that describe the overall shape of the binned region: \{$R_{\rm s}$, $q_{\rm s}$, $\phi_{\rm s}$\}.  

In order to generate the datacube, we effectively generate an image at every wavelength interval, with the source-plane flux at a given wavelength being determined by the $v_{\rm LOS}$, $\sigma_{v}$, and surface-brightness profiles.  I.e., we are producing a source light profile based the source profiles for a given wavelength, and as strong lensing is achromatic in nature, we ray-trace and produce a single, spectral-resolution-convolved wavelength slice; this process is repeated across each wavelength in our datacube while keeping the lens model fixed.  We assume a functional form of the [\ion{O}{II}] doublet with two Gaussian profiles (at restframe 3727.09\AA\ and 3729.88\AA).  The line ratio of the doublet is fixed to 1.30, which is the measured line ratio from a large aperture (arbitrary defined around each lensed image to encompass nearly all of [\ion{O}{II}] emission) fit of the spectra across all images; qualitatively, we do not observe that the line ratio changes significantly across images.  

All centroids of the three profiles describing S4 (\{$c_{\rm s, x}$, $c_{\rm s, y}$\}) are jointly constrained.  And from our datacube reconstruction, we fit our model to the observed continuum-subtracted datacube described in Section~\ref{sec:muse}.  In total, we sample and constrain 20 parameters: eight parameters describing the convergence profile \{$\theta_{\rm E}$, $\gamma$, $q_{\rm m}$, $\phi_{\rm m}$, $c_{\rm x}$, $c_{\rm y}$, $\gamma_{\rm ext}$, $\phi_{\rm ext}$\}, four describing the $v_{\rm LOS}$ profile \{$A_{v}$, $R_{v}$, $C_{v}$, ${\phi}_{v}$\}, three describing the $\sigma_{v}$ profile \{$A_\sigma$, $R_\sigma$, $C_\sigma$\}, and five describing the surface brightness and shape of S4 \{$R_{\rm s}$, $q_{\rm s}$, $\phi_{\rm s}$, $c_{\rm s, x}$, $c_{\rm s, y}$\}.  We employ broad uniform priors for all variables as to not bias the result; we find that all resulting posteriors are not limited by these bounds.  Much like the 2D image modeling in Section~\ref{subsec:2dmodeling}, we run an initial PSO algorithm to explore the parameter space, followed by an \texttt{EMCEE} MCMC algorithm (run until all the chains have converged) to probe the posteriors of these parameters \citep{emcee}.  

We have packaged our flexible framework used for this project into the Python package \texttt{IFUL}\footnote{\url{https://github.com/williyamshoe/iful}} (Integral Field Unit Lensing; Sheu~et~al. in preparation).

\section{Results} \label{sec:results}
\begin{figure}
\begin{center}
 \includegraphics[width=1\linewidth]{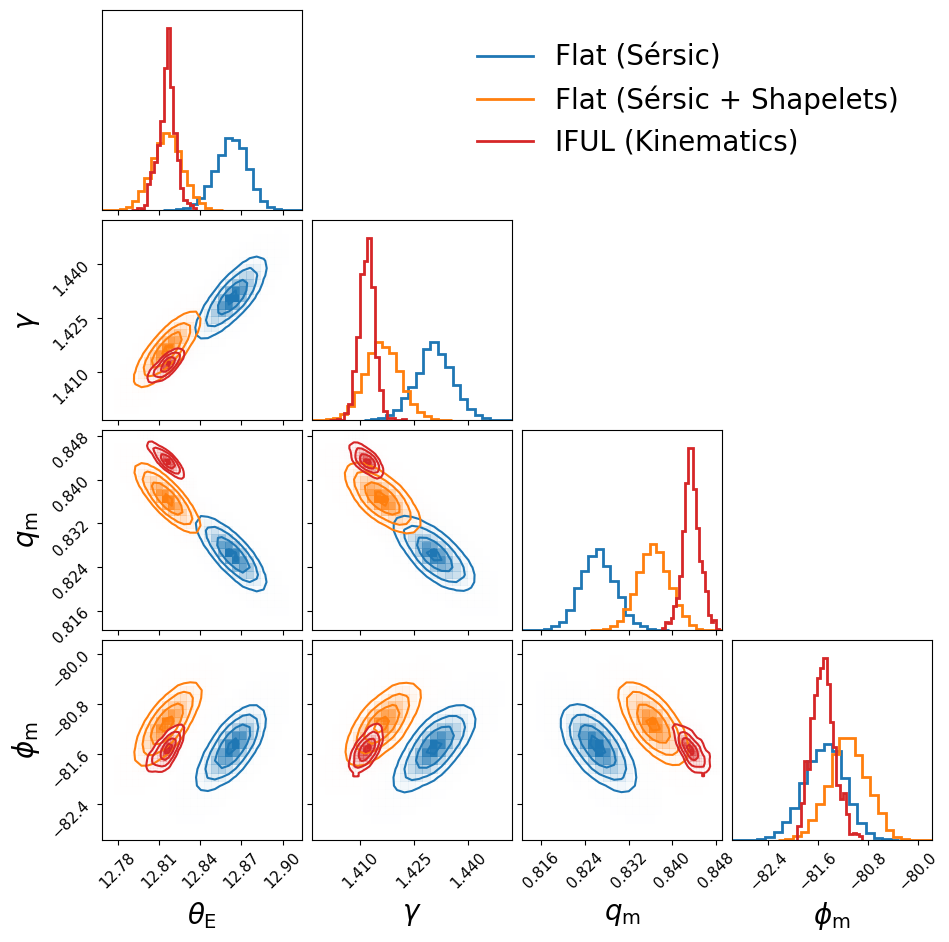}
 \caption{Corner plot comparison between three different modeling approaches (see Table~\ref{tab:tam_comp} and Section~\ref{subsec:model_comparision}).  We show the Einstein radius $\theta_{\rm E}$, logarithmic slope $\gamma$, the semi-major to minor axis ratio $q_{\rm m}$, and the position angle $\phi_{\rm m}$ of the power-law profile describing the primary deflector convergence profile.}
 \label{fig:corner_comp}
\end{center}
\end{figure}

\begin{table}
\renewcommand{\arraystretch}{1.25}
\begin{center}
\caption{The resulting posterior distributions for the source kinematic parameters (both $v_{\rm LOS}$ and $\sigma_v$ profiles) of our datacube-based model.  $\phi_{v}$ is given as East of North.  See Equations~\ref{eq:vlos} and \ref{eq:sigmav} for the analytical context regarding these variables.  }\label{tab:v_params}
\begin{tabular}{cc|c}
\hline
 Parameter & Units & Posterior \\
\hline
$A_{v}$ & [km s$^{-1}$] & $266.2^{+11.3}_{-9.9}$\\
$R_{v}$ & [$''$] & $0.246^{+0.016}_{-0.015}$\\
$C_{v} - 1.432c$ & [km s$^{-1}$] & $61.44^{+0.85}_{-0.97}$\\
$\phi_{v}$ & [$^{\circ}$] & $93.86^{+0.48}_{-0.49}$\\
$A_{\sigma}$ & [km s$^{-1}$] & $132.3^{+38.7}_{-31.1}$\\
$R_{\sigma}$ & [$''$] & $0.017^{+0.003}_{-0.002}$\\
$C_{\sigma}$ & [km s$^{-1}$] & $37.16^{+0.38}_{-0.38}$\\
\hline
\end{tabular}
\end{center}
\end{table}

\begin{figure*}
\begin{center}
 \includegraphics[width=0.95\linewidth]{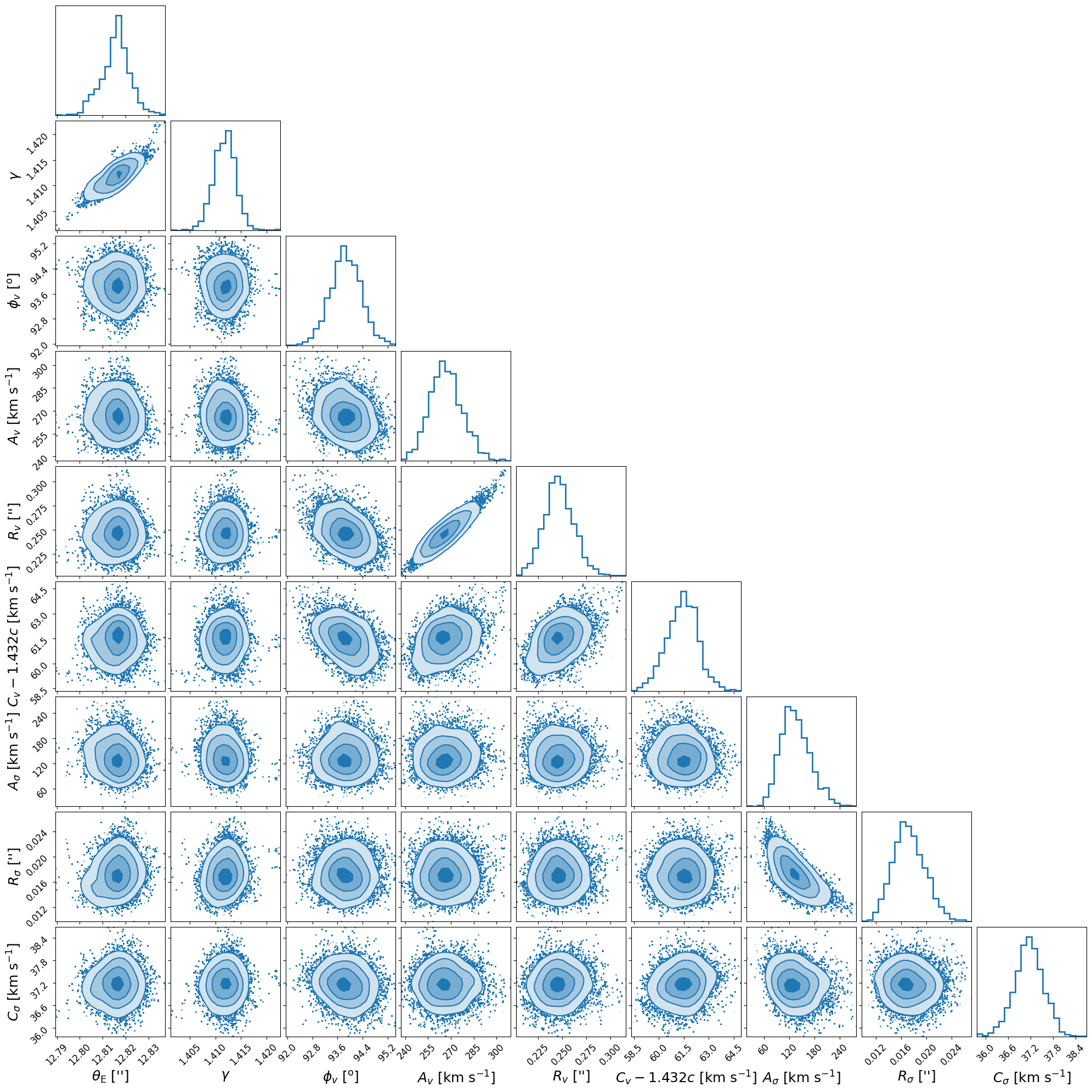}
 \caption{Corner plot for a subset of parameters constrained with our \texttt{IFUL}, datacube-informed model.  $\phi_{v}$ is given as East of North.  All parameters shown are fairly Gaussian in shape.}
 \label{fig:corner}
\end{center}
\end{figure*}

Here we present our results from our datacube-based modeling methodology.  In Figure~\ref{fig:model_wave}, we show various wavelength slices of the continuum-subtracted datacube, the modeled datacube, and the resulting residuals of our best-fitting parameterization.  Figure~\ref{fig:rgb_comp} illustrates an alternative RGB-colored image representation of the observed and modeled datacube.  The lens model posteriors are given in Table~\ref{tab:tam_comp} and Figure~\ref{fig:corner_comp}, along with posteriors of other comparison models.  In Figure~\ref{fig:source_kinematics}, we present the best-fitting model's kinematic profiles of S4, with the posterior values given in Table~\ref{tab:v_params}.  And finally in Figure~\ref{fig:corner}, we display a corner plot of a subset of parameters constrained with our \texttt{IFUL} model.

\subsection{Model comparison}\label{subsec:model_comparision}
We run two flat image-based models to compare our datacube-based modeling results.  The first model recycles the flat model using a single Sérsic profile to represent the source galaxy light, which was used as an initial parameterization for our \texttt{IFUL} model (see Section~\ref{subsec:2dmodeling}).  The second flat model expands on this, where we now use a single Sérsic profile with an additional shapelets basis (with a maximum Hermite polynomial order of $n_{\rm max} = 4$) for the source light \citep{shapelets1, shapelets2}.  

As seen in Figure~\ref{fig:corner_comp} and Table~\ref{tab:tam_comp}, our datacube modeling scheme was able to achieve significantly tighter statistical uncertainties on lensing parameters, especially on $\theta_{\rm E}$ and $\gamma$.  Relative to the image-only shapelets model fit to the same data, the uncertainties on the convergence parameters shrink by roughly a factor of two to three: a factor of $1.8$ on $\theta_{\rm E}$, $2.5$ on $\gamma$, and $2.0$ on $q_{\rm m}$.  We are successfully able to extract extra lensing constraints from forward modeling the source galaxy kinematic profile.  

Also worth noting is that the best fit parameters shift in the same direction when including a more complicated source light model (for both blue to orange, and blue to red in Figure ~\ref{fig:corner_comp}).  This may suggest that the choice of source-light model could bias the lens parameters by more than statistical uncertainties.  Though we find that both a $n_{\rm max} = 4$ shapelets model and a binned \texttt{IFUL} model individually well-captures the source light complexities, a more comprehensive investigation of source-light model and its impact on the inferred lens parameters will be left to a future study.

Compared to the posteriors achieved in Papers~\citetalias{Sheu_2024} and \citetalias{urcelay2026}, our uncertainties are much larger and in tension: Paper~\citetalias{Sheu_2024} predicted $\theta_{\rm E} = 13\farcs03 \pm 0.02$ and $\gamma = 1.67 \pm 0.01$, whereas Paper~\citetalias{urcelay2026} predicted $\theta_{\rm E} = 12\farcs738 \pm 0.005 $ and $\gamma = 1.571 \pm 0.001$.  While the higher resolution \textit{HST} imaging does partially explain this, we also believe that it is because these papers not only have different modeling assumptions, but also constrain the primary power-law profile across multiple source redshifts whereas we only focus on S4 at $z_{\rm s}=1.432$.  Furthermore, as we are also probing near the center of the BCG with the central image $4e$, our model is probing the smaller radii compared to the larger arcs modeled in Papers~\citetalias{Sheu_2024} and \citetalias{urcelay2026}, which may explain our notably shallower $\gamma$ measurement.  This may indicate that the Carousel Lens necessitates a radius-dependent convergence profile, but additional studies at larger radii are necessary to definitively prove this. 

This analysis nevertheless serves as a proof-of-concept that it is possible and practical to utilize the source kinematic information to improve lens models.  In this current regime of systematics-dominated lens models, this methodology serves as an alternative and complementary modeling scheme to better understand our modeling systematics.  Rather than solely relying on high-resolution imaging data, this approach can provide a systematic check using completely different and independent data.  Additionally, we are now able to probe smaller radii due to the removal of the lens light and the reveal of hidden demagnified images.

\subsection{Source profiles}
From the best-fitting parametrization in Figure~\ref{fig:source_kinematics}, we present the unlensed $v_{\rm LOS}$, $\sigma_v$, and surface-brightness profiles of S4, with the full posteriors in Table~\ref{tab:v_params}.  From the $v_{\rm LOS}$ profile, we can see that the model captures the clear and strong line-of-sight rotation in S4 that we expect from the observed datacube (seen in Figure~\ref{fig:wave}), with an asymptotic rotation velocity of $A_{v} = 266.2^{+11.3}_{-9.9}$~km~s$^{-1}$.  The discretized and linearly-fitted light profile is also within expectation, with a brighter inner region that gradually fades at larger radii.  The bin sizes adequately capture the complexity of the light profile.  We also note that the profile presented is the unconvolved profile; when convolved with the PSF, the discretized character of the profile disappears and it appears smooth.  Our LOS velocity profile captures the rotation we observe from the lensed images, but now characterized at the source plane with an asymptotic rotational velocity of $A_{v} = 266.2^{+11.3}_{-9.9}$ km s$^{-1}$ and a rotational scale radius of $R_{v} = 0\farcs246^{+0.016}_{-0.015}$.  The velocity dispersion profile is also well constrained, with a bulge velocity dispersion $A_{\sigma}=132.3^{+38.7}_{-31.1}$ km s$^{-1}$, dispersion scale radius $R_{\sigma}=0\farcs017^{+0.003}_{-0.002}$, and a velocity dispersion floor $C_{\sigma}=37.16^{+0.38}_{-0.38}$ km s$^{-1}$.  As the posterior of $R_{\sigma}$ is Gaussian-like and does not hug the $R_{\sigma}=0''$ floor (Figure~\ref{fig:corner}), this indicates that we are able to resolve the galaxy bulge from the velocity dispersion through marginalizing all lensed images in our macro model.  

\begin{figure}
\begin{center}
 \includegraphics[width=1\linewidth]{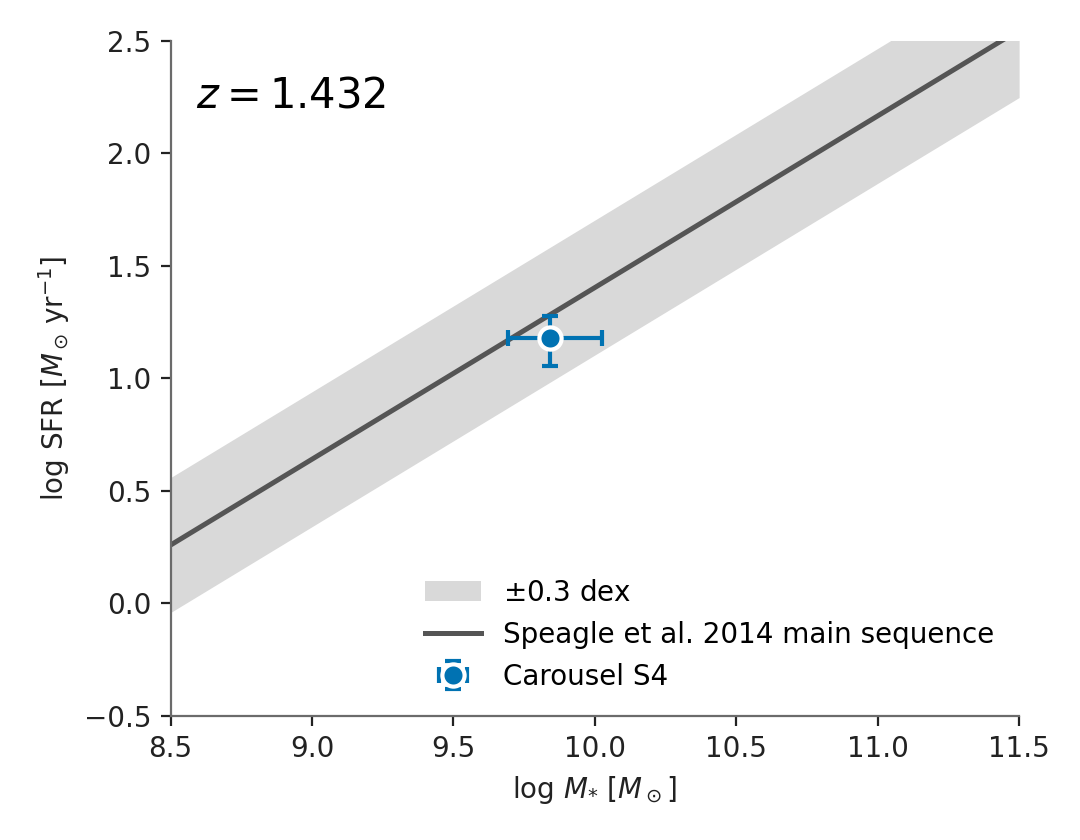}
 \caption{The star-forming galaxy main sequence track of the stellar mass versus star formation rate, at $z=1.432$ \citep[derived from Equation~30 in][]{Speagle2014}.  We estimate the stellar mass of S4 by using our dynamical mass measurement and a conversion factor of $\log(M_{*} /M_{\rm dyn}) = -0.51 \pm 0.11$ \citep[from][]{wyrst}.  We find that S4 lies confidently along the SFG main sequence.}
 \label{fig:sfg_ms}
\end{center}
\end{figure}

Additionally, we are also able to estimate the SMBH mass through the $M_{\rm BH}$ $-$ circular velocity relation \citep{sun2013}:
\begin{equation}
    \log_{10}\left(\frac{M_{BH}}{M_\odot}\right) = 7.43 + 3.68 \log\left(\frac{A_v}{200\text{ km s}^{-1}}\right),
\end{equation}
measuring $\log M_{\rm BH}/M_{\odot} = 7.89 \pm 0.51$ (intrinsic scatter from \citealp{sun2013}).  This is in line with .  A similar analysis of an inactive, high-$z$ black hole was presented in \citet{Newman_2026}, which uses \textit{James Webb Space Telescope} (\textit{JWST}) NIRSpec IFU observations of a strongly lensed source at $z\approx 1.95$ to constrain the mass of its central SMBH.  However, our approach differs in a few aspects; e.g., we forward model the macro-lens model whereas the lens model is fixed in \citet{Newman_2026}.  Their analysis also includes a Jeans anisotropic modeling of the lensed source galaxy \citep{Cappellari2008, Cappellari2020, Cappellari2026}, whereas we parametrically map out the kinematics of the source plane and rely on the $M_{\rm BH}$ $-$ circular velocity relation for our measurement. Additionally, the analysis in \cite{Newman_2026} uses a single lensed image, whereas we simultaneously model all observed images of S4. %Due to computational limitations, we are unable to build a dynamical model in conjunction with the lens modeling scheme, and as such our $\log M_{\rm BH} / M_{\odot}$ measurement may require additional systematic checks or dynamical modeling for further confirmation.

From our model of S4, evidence seems to point that it is a typical star-forming galaxy (SFG).  From the surface-brightness profile, we measure a circular effective radius of $R_{\rm eff} = 0\farcs282 \pm 0.002$.  Assuming an intrinsic scale height to scale length thickness of $q_0 = 0.25$ \citep{wyrst}, we measure an inclination angle of $i=47.9^\circ \pm 8.5^\circ$.  We can then calculate the dynamical mass with:
\begin{equation}
    M_{\rm dyn}(<r) = r\frac{v_{\rm circ}^2(r)}{G},
\end{equation}
where the pressure-supported, circular velocity is
\begin{equation}
    v^2_{\rm circ}(r) = \left(\frac{v_{\rm LOS}(r)}{\sin i}\right)^2  + 2C_{\sigma}^2.
\end{equation}
We measure a dynamical mass of $\log M_{\rm dyn}(<R_{\rm e}) / M_{\odot} = 10.35^{+0.15}_{-0.10}$.  Finally, we can estimate the star formation rate SFR from the [\ion{O}{II}] emission flux using the relation from \citet{kewley}, where we find SFR[\ion{O}{II}] $= 15.1 \pm 3.8\ M_{\odot}{\rm\ yr}^{-1}$.  These measurements are all within expectation for a late-type galaxy at $z_{\rm s} = 1.432$ along the SFG main sequence (see Figure~\ref{fig:sfg_ms}).

\section{Summary and conclusions} \label{sec:conclusion}
In this analysis, we present a novel datacube-based modeling scheme for lens modeling, incorporating both spatial and spectral dimensions to provide additional lensing constraints.  We have measured an Einstein radius of $\theta_{\rm E} = 12\farcs816^{+0.005}_{-0.007}$ and a logarithmic slope of $\gamma = 1.412^{+0.002}_{-0.002}$ from S4's images at $z_{\rm s}=1.432$.  As the Carousel Lens is exceptional in being a relaxed cluster-scale lens with $\ge10$ lensed sources, constraints on $\theta_{\rm E}$ are especially important and propagate directly to the $w$ and $\Omega_{\rm m, 0}$ cosmography uncertainties.  

We also demonstrate the feasibility of this modeling framework by recovering tighter statistical constraints on lensing compared to image-only control models.  From this forward-modeling approach, we gain a stronger understanding of the source galaxy's kinematics by marginalizing across the lensed images, with constraints on the line-of-sight rotation profile and the mass of the residing SMBH of $\log M_{\rm BH} / M_{\odot} = 7.89 \pm 0.51$.  We also measure a dynamical mass of $\log M_{\rm dyn}(<R_{\rm e}) / M_{\odot} = 10.35^{+0.15}_{-0.10}$ and SFR[\ion{O}{II}] $= 15.1 \pm 3.8\ M_{\odot}{\rm\ yr}^{-1}$, which lies squarely along the SFG main sequence track.  While further studies to probe systematics are necessary, this can give unparalleled resolution and insight into the kinematic structure of high-redshift late-type galaxies.  In conjunction with compound lensing, kinematic constraints on the intermediate source galaxies allow future studies to account for the internal structure of S4, constraining the multi-plane mass sheet degeneracy \citep{Schneider_2019, teodori2026}.

As this methodology is early in its development, improvements can and should be made for future studies.  As stated previously in Section~\ref{sec:muse}, we conservatively overestimate the per-spaxel variance to compensate for the underestimated covariance from the MUSE data reduction pipeline.  Hence, a better understanding of the covariance between the VLT/MUSE IFU spaxels would greatly improve our model.  In addition, further tests using other source redshifts of the Carousel Lens, together with more flexible velocity profiles, would better probe the systematics of our model. 

Utilizing this technique on high-resolution \textit{JWST}/NIRSpec IFU data of the Carousel Lens arcs would provide even tighter constraints on lens models and deeper insight into the source galaxy kinematics.  Moreover, high-resolution IFU spectroscopy has already been obtained for other strong lensing systems \citep[e.g.,][]{tdcosmo2025, sheu1433, knabel2026}.  Our pipeline can be used to generate additional independent lens models for these and other IFU-observed systems, which is important since lens modeling has already advanced to the point where systematics are the main driver for uncertainty.  

\begin{acknowledgements}
This research used resources of the National Energy Research Scientific Computing Center (NERSC), a U.S. Department of Energy Office of Science User Facility operated under Contract No. DE-AC02-05CH11231, and the Computational HEP program in The Department of Energy's Science Office of High Energy Physics provided resources through the ``Cosmology Data Repository'' project (Grant \#KA2401022).
\end{acknowledgements}

% \appendix
% \section{Append A}
% \label{app:A}

\bibliographystyle{aa}
\bibliography{biblio}

\end{document}